\documentclass[tikz,11pt,english,a4paper]{article}
\usepackage{float}
\usepackage{jcappub}
\usepackage[utf8]{inputenc}

\pdfoutput=1

\usepackage{fancyvrb}
\usepackage{keyval}

\usepackage{bm}
\usepackage{algorithm}
\usepackage{nicefrac}
\usepackage{bbold}
\usepackage{dsfont}
\usepackage{setspace}
\usepackage[linewidth=1pt]{mdframed}
\usepackage{bold-extra}
\usepackage{tabto}

\usepackage{algorithm}
\usepackage[noend]{algpseudocode}
\makeatletter
\def\BState{\State\hskip-\ALG@thistlm}
\makeatother

\usepackage{fontawesome5}
\usepackage{hyperref}
\makeatletter
\newcommand{\github}[1]{%
   \href{#1}{\faGithub}%
}
\makeatother

\newcommand{\connect}{\textsc{connect}}
\newcommand{\class}{\textsc{class}}
\newcommand{\camb}{\textsc{camb}}
\newcommand{\client}{\textsc{cl}\text{i}\textsc{ent}}
\newcommand{\emcee}{\textsc{emcee}}
\newcommand{\montepython}{\textsc{MontePython}}
\newcommand{\tf}{TensorFlow}
\newcommand{\cobaya}{Cobaya}
\usepackage{siunitx}
\DeclareSIUnit \parsec {pc}

\usepackage[T1]{fontenc}

\usepackage{lmodern} 
\rmfamily 
\DeclareFontShape{T1}{lmr}{b}{sc}{<->ssub*cmr/bx/sc}{}
\DeclareFontShape{T1}{lmr}{bx}{sc}{<->ssub*cmr/bx/sc}{}

\usepackage{soul}

\usepackage[edges]{forest}
\definecolor{folderbg}{RGB}{124,166,198}
\definecolor{folderborder}{RGB}{110,144,169}
\newlength\Size
\tikzset{%
  folder/.pic={%
    \filldraw [draw=folderborder, top color=folderbg!50, bottom color=folderbg] (-1.05*\Size,0.2\Size+5pt) rectangle ++(.75*\Size,-0.2\Size-5pt);
    \filldraw [draw=folderborder, top color=folderbg!50, bottom color=folderbg] (-1.15*\Size,-\Size) rectangle (1.15*\Size,\Size);
  },
  file/.pic={%
    \filldraw [draw=folderborder, top color=folderbg!5, bottom color=folderbg!10] (-\Size,.4*\Size+5pt) coordinate (a) |- (\Size,-1.2*\Size) coordinate (b) -- ++(0,1.6*\Size) coordinate (c) -- ++(-5pt,5pt) coordinate (d) -- cycle (d) |- (c) ;
  },
}
\forestset{%
  declare autowrapped toks={pic me}{},
  declare boolean register={pic root},
  pic root=0,
  pic dir tree/.style={%
    for tree={%
      folder,
      font=\ttfamily,
      grow'=0,
    },
    before typesetting nodes={%
      for tree={%
        edge label+/.option={pic me},
      },
      if pic root={
        tikz+={
          \pic at ([xshift=\Size].west) {folder};
        },
        align={l}
      }{},
    },
  },
  pic me set/.code n args=2{%
    \forestset{%
      #1/.style={%
        inner xsep=2\Size,
        pic me={pic {#2}},
      }
    }
  },
  pic me set={directory}{folder},
  pic me set={file}{file},
}

\usepackage{newfloat}
\usepackage{caption}
\DeclareFloatingEnvironment[fileext=frm,placement={!t},name=Algorithm]{pseudoenv}
\DeclareCaptionLabelSeparator{normal-colon}{\normalfont : } 
\makeatletter
\g@addto@macro\bfseries{\boldmath}
\makeatother

\makeatletter
\def\old@comma{,}
\catcode`\,=13
\def,{%
  \ifmmode%
    \old@comma\discretionary{}{}{}%
  \else%
    \old@comma%
  \fi%
}
\makeatother

\newcommand{\numStdVar}{\kappa}
\newcommand{\VarText}{kappa}               

\begin{document}


\title{{\rmfamily\textsc{CL}i\textsc{ENT}}: A new tool for emulating cosmological likelihoods using deep neural networks}

\author[a]{Luca Janken,}
\author[a]{Steen Hannestad, }
\author[a]{Thomas Tram,}
\author[b]{Andreas Nygaard}

\affiliation[a]{Department of Physics and Astronomy, Aarhus University,
 DK-8000 Aarhus C, Denmark}
\affiliation[b]{Department of Astrophysics, University of Zurich,
 CH-8057 Zurich, Switzerland}

\emailAdd{luca.janken@post.au.dk}
\emailAdd{steen@phys.au.dk}
\emailAdd{thomas.tram@phys.au.dk}
\emailAdd{andreas@phys.au.dk}
\emailAdd{andreas.hansen@uzh.ch}

\abstract{
	Cosmological emulation of observables such as the Cosmic Microwave Background (CMB) spectra and matter power spectra have become increasingly common in recent years because of the potential for saving computation time in connection with cosmological parameter inference or model comparison. In this paper we present \client{} (Cosmological Likelihood Emulator using Neural networks with \tf), a new method which circumvents the computation of observables in favour of directly emulating the likelihood function for a data set given a model parameter vector. We find that the method is competitive with observable emulators in terms of the required number of function evaluations, but has the distinct advantage of producing a surrogate likelihood which is completely auto-differentiable.
Using less than $2 \times 10^4$ function evaluations \client{} typically achieves credible intervals within better than $0.1 \sigma$ of those obtained using the true likelihood and single-point emulator precision better than $\Delta \chi^2 \sim 0.5$ across relevant regions in parameter space.
		}

\maketitle

\section{Introduction}\label{sec:introduction}

The traditional pipeline for performing cosmological Bayesian parameter inference or model comparison involves the computation of one or more observables such as the Cosmic Microwave Background (CMB) spectrum or the large scale matter power spectrum for a very large number of different model parameters - often requiring millions of model computations (see e.g.\ \cite{Audren:2012wb,Torrado:2020dgo}).
For CMB analysis a single point requires running an Einstein--Boltzmann solver such as \camb{} \cite{Lewis:1999bs} or \class{} \cite{Blas:2011rf} to compute the CMB power spectrum, followed by the evaluation of the likelihood on this particular point. The CPU time requirement is typically tens of core seconds for a single point, making
parameter inference computationally expensive and Bayesian model comparison prohibitively expensive.
Analysing large scale structure in the non-linear regime carries a much higher cost because it requires running an $N$-body simulation to compute the power spectrum, de facto making a brute force approach to Bayesian model comparison impossible.

This problem has made emulation of observables increasingly popular in cosmology and a number of different frameworks are now available for this purpose
(see e.g.\  \cite{SpurioMancini:2021ppk,Nygaard:2022wri,Gunther:2025xrq,Auld:2006pm,Fendt:2006uh,Lazanu:2025xdk,Heitmann:2013bra,Euclid:2018mlb,Arico:2021izc,Moran:2022iwe,Bonici:2023cyn}). However, since emulation of cosmological observables still requires a subsequent direct computation of the likelihood, parameter inference or model comparison can still be very expensive in terms of computational resources in the event that the likelihood is complex. Furthermore, while the emulator possesses the desirable feature of being auto-differentiable, this property is typically lost in the likelihood computation, although differentiable likelihood codes such as \texttt{candl}~\cite{Balkenhol:2024candl}, \texttt{clipy}~\cite{clipy}, and \texttt{desilike}~\cite{desilike} are starting to emerge.
This problem makes the direct emulation of the likelihood function from the model parameter vector a desirable goal. A related but distinct approach is implicit likelihood inference (ILI, also referred to as likelihood-free inference or simulation-based inference), where the data-generating forward model is emulated rather than the likelihood itself. While ILI bypasses the need for an explicit likelihood computation, it requires direct access to the data or carefully chosen summary statistics, rather than pre-compressed likelihood evaluations.

Here, we present \client{}, a general framework for emulating likelihood functions using \tf\ \cite{Abadi:2016kic}. The current version allows for the emulation of any likelihood currently supported by 
the industry standard Markov Chain Monte Carlo (MCMC) samplers
\montepython{}  \cite{Audren:2012wb} or  \cobaya{} \cite{Torrado:2020dgo}.
The algorithm works by iteratively training the network on a cloud of points gathered so that their density asymptotes to follow a tempered version of the actual likelihood function
\footnote{Note that the method presented here bears some similarities to the tempered gathering of training data described in \cite{Negri:2025cyc}  for the purpose of emulating gravitational wave signals.}. This allows it to gather training data which maximises knowledge about the target likelihood function while minimising the total computational cost. A similar active learning approach is utilised in the GPry code~\cite{Gammal:2022eob}, where the likelihood is emulated using a Gaussian Process Regressor (GPR). In this framework, new training data are iteratively selected via an acquisition function designed to reduce emulator uncertainty across the parameter space. The trained GPR is then used for inference in combination with standard Metropolis–Hastings and nested samplers. While this approach is highly effective in low-dimensional settings, its computational cost scales poorly with dimensionality, making applications beyond $\sim\!\!15$ parameters increasingly challenging in practice. In addition, the resulting GPR surrogate, as implemented in GPry, is not readily compatible with gradient-based inference methods that rely on efficient auto-differentiation. While GPry is designed to accelerate or replace the sampling procedure itself, \client{} instead focuses on constructing a high-fidelity likelihood emulator as the primary end product. As such, the two approaches are tailored to distinct, and in many cases complementary, use cases.

In Section~\ref{sec:lik-fun} we outline the likelihood functions used in the current work, and in Section~\ref{sec:framework} we provide details of the emulator framework. Section~\ref{sec:metrics} contains a discussion of the metrics we use for benchmarking the emulator performance, and Section~\ref{sec:convergence} describes the convergence criterion for the training procedure. Section~\ref{sec:results} contains numerical results for both synthetic and actual likelihoods, and finally, in Section~\ref{sec:conclusion} we provide our conclusions.

\section{Likelihood functions used}\label{sec:lik-fun}

Our ultimate aim is to emulate the likelihood functions typically used in cosmological data analysis, and the most complex likelihood function pipeline currently used is that of the CMB anisotropies measured by the Planck mission \cite{Planck:2019nip}.
Depending on the specific implementation, this likelihood function depends on all cosmological model parameters (i.e.\ 6 in the standard $\Lambda$CDM model: the physical baryon density $\Omega_{\rm b} h^2$, the physical cold dark matter density $\Omega_{\rm cdm} h^2$, the angular size of the sound horizon $100\theta_{\rm s}$, the amplitude of primordial scalar perturbations $\ln(10^{10} A_{\rm s})$, the scalar spectral index $n_{\rm s}$, and the optical depth to reionisation $\tau_{\rm reio}$) plus of order 20 experimental nuisance parameters. For the analysis presented here we will use TT,TE,EE+lowE+lensing CMB data from Planck 2018~\cite{Planck:2019nip} in combination with BAO measurements from BOSS DR12~\cite{boss2016}, 6dF~\cite{Beutler:2011hx} and MGS~\cite{ross2014}, corresponding to \texttt{base2018TTTEEE\_lensing\_bao} in \montepython{}. In the remainder of this paper we will refer to this likelihood combination simply as the ``Planck likelihood''.

Using this configuration, each single likelihood evaluation then consists of a call to \class{} taking approximately 10 core-seconds for the simplest models plus a call to the likelihood which takes of order 200 ms on a single core. This typically means several hours or more of wall time consumption for each test performed.

In order to speed up testing of the code we instead perform all our initial tests using a synthetic and fully analytic likelihood function. The simplest choice for this function is 
a multi-variate Gaussian of dimension $N=N_{\rm cosmology} + N_{\rm nuisance}$ using the same parameter covariance matrix, $\mathbf{C}$, as the actual Planck likelihood.
\begin{equation}
\mathcal{L}(\boldsymbol{\theta}) =
\frac{1}{\sqrt{(2\pi)^N \det \mathbf{C}}}
\exp\!\left[
-\frac{1}{2}
(\boldsymbol{\theta} - \boldsymbol{\mu})^{\mathrm{T}}
\mathbf{C}^{-1}
(\boldsymbol{\theta} - \boldsymbol{\mu})
\right].
\end{equation}

In order to also test a non-Gaussian yet simple analytic likelihood function we have extended this setup with 2 additional parameters which are uncorrelated with all other parameters and for which the contribution to the likelihood is of the form $e^{-\theta_{N+1} \theta_{N+2}}$ 
An example of such a likelihood could be the presence of an additional sterile neutrino of mass $m_s$ and effective number $N_{\text{eff},s}$ so that the total contribution to the energy density at late times is $\rho \propto m_s N_{\text{eff},s}$. However, note that the two non-Gaussian parameter directions do not correspond to the actual sterile neutrino model we use as one of the real data test cases so that inferred parameter values for this case do not correspond to the sterile neutrino parameters estimated from Planck data. For the analytic likelihood we will refer to these two non-Gaussian parameters simply as $x_{28}$ and $x_{29}$.
The similarity of this analytic function to that of the actual Planck likelihood makes it very efficient for tuning hyperparameters of both the network and the training procedure, and we will see later that the network and training structure perform very well on actual Planck data.
This means that in total we use four different likelihoods to test the \client{} framework:

\begin{itemize}

\item A 27-dimensional analytic Gaussian with covariance set to be identical to that of the actual Planck data as described above.

\item A 29-dimensional analytic likelihood with two new parameters added which are uncorrelated with all other parameters, but for which the contribution to the likelihood is non-Gaussian.

\item The 27-dimensional Planck likelihood for $\Lambda$CDM, corresponding to $\Omega_{\rm b} h^2, \Omega_{\rm cdm} h^2, 100 \theta_{\rm s}, \ln(10^{10} A_{\rm s}), n_{\rm s},  \tau_{\rm reio}$ as cosmological parameters with an additional 21 nuisance parameters.

\item The 29-dimensional Planck likelihood for $\Lambda$CDM + $(N_{\text{eff},s},m_s)$

\end{itemize}


\section{Emulator framework}\label{sec:framework}

\subsection{Loss function}\label{sec:loss}

Emulation of the likelihood function differs somewhat from general function emulation in the sense that use of the emulated function typically comes in the form of Bayesian parameter inference, maximum likelihood analysis, or Bayesian evidence calculations. All three methods are sensitive to the exact value of the likelihood function in regions of large likelihood, but insensitive in regions of moderate or low likelihood. 
This makes the use of a traditional mean squared error loss non-optimal. Rather, it seems more appropriate to use a loss function based on a relative error

\begin{equation}
{\rm loss} \propto \left\langle \left( \frac{\chi^2_{\rm surrogate} - \chi^2_{\rm exact}}{\chi^2_{\rm exact} }\right)^2 \right\rangle \,, \label{eq:loss1}
\end{equation}
where we have defined an effective $\chi^2$ in terms of the likelihood as
\begin{equation}
\chi^2 \equiv -2 \left(\log(\mathcal{L}) - \log(\mathcal{L}_\text{best}) \right) \,,
\end{equation}
and $\log(\mathcal{L}_\text{best})$ is the largest value of $\mathcal{L}$ discovered so far. However, the loss-function in equation~\eqref{eq:loss1} would diverge as $\chi^2_{\rm exact} \rightarrow \chi^2_{\rm best}=0$ (note that by our definition, $\chi^2$ approaches zero at the best-fit point, irrespective of whether we are dealing with synthetic or real data). We therefore regularise it in the following way,
\begin{equation}
{\rm loss} \propto \left\langle \left( \frac{\chi^2_{\rm surrogate} - \chi^2_{\rm exact}}{\chi^2_{\rm exact} + \epsilon}\right)^2 \right\rangle \,, \label{eq:loss2}
\end{equation}
where $\epsilon>0$. For points where $\chi^2_{\rm exact} \ll \epsilon$, the loss reduces to the standard mean squared error, whereas it becomes a relative error whenever $\chi^2_{\rm exact} \gg \epsilon$. Thus, $\epsilon$ decides how far away from the best-fit a point should be before it is penalised. This $\epsilon$ parameter will depend on the dimensions of the problem, so instead we define a range based on the number of standard deviations away from the best-fit. Using the well-known Wilson–Hilferty transformation~\cite{wilson1931distribution} we find the formula

\begin{equation}
\epsilon = \Delta \chi^2 \simeq n \left(1-\frac{2}{9n}+\numStdVar\sqrt{\frac{2}{9n}} \right)^3\,,
\end{equation}

where $\numStdVar$ is the number of standard deviations and $n$ is the number of dimensions. (Additional details can be found in appendix~\ref{sec:quantile}). A priori, one would think that $1 \lesssim \numStdVar \lesssim 10$ would be reasonable.

\begin{figure}[tb]
    \centering
    \includegraphics[width=\textwidth]{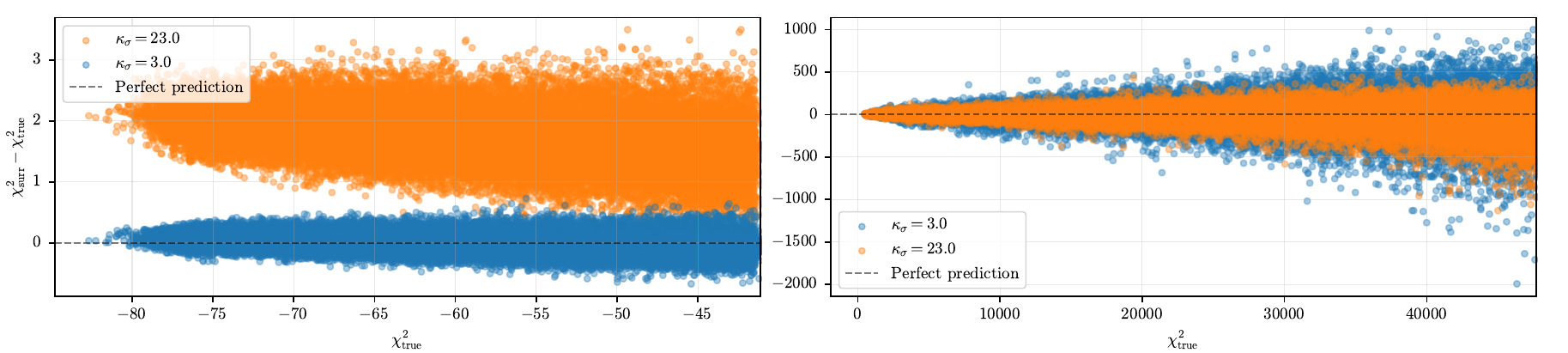}
    \caption{Surrogate error as a function of exact $\chi^2$ value for networks trained on the 27-dimensional analytic Gaussian likelihood. The left panel shows the region closest to the best-fit while the right panel shows the error deep in the tails. A lower value of the parameter $\numStdVar$ in the loss function clearly reduce the scatter of the emulated likelihood  around the best-fit at the expense of increasing the scatter further away.}
    \label{fig:surrogate_error}
\end{figure}

Figure \ref{fig:surrogate_error} shows the surrogate error evaluated on a cloud of points with varying $\chi^2$ for two different values of $\numStdVar$. As expected, a small value of $\numStdVar$ pushes the network to emulate points close to the likelihood maximum very well, but at the price of larger uncertainties in the tails. It should also be noted that the network trained using a large value of $\numStdVar$ exhibits a systematic bias at small $\chi^2$.

In the current work the main focus will be on emulating the likelihood well enough to extract accurate posteriors, and we will therefore use $\numStdVar=3$ for our actual runs. If the main goal was profile likelihoods an even smaller value of $\numStdVar$ might be favourable, whereas if the goal was to calculate the Bayesian evidence it might be better to use a larger value in order to be more accurate in the tails. 

\subsection{Activation function}\label{sec:activation}

In a neural network, an activation function is a non-linear function applied element-wise to the output of each neuron, allowing the network to learn complex, non-linear mappings between inputs and outputs. Without activation functions, a deep network would reduce to a single linear transformation, regardless of its depth.

The emulation problem discussed here consists of mapping a set of real model (and nuisance) parameters to a real-valued function which has a minimum at some finite value, and which diverges quadratically for values of model parameters far removed from the best-fit point.
Such a problem is typically not well handled by activation functions which are bounded, i.e. $\tanh$ or similar, meaning that it is typically better to use e.g.\ a ReLU or similar. However, as was also discussed in \cite{Nygaard:2022wri}, ReLU is also not an ideal choice because it lacks expressivity at negative values. Instead we have settled on the same choice of activation function as in the \connect{} framework, namely the Alsing et al. activation function \cite{Alsing:2020} which replaces ReLU with a linear function at negative values and is characterised by two hyperparameters: The asymptotic slope, $\gamma$, of the linear function at negative values and the broadness, $\beta$, of the transition region between the asymptotic functions at negative and positive values. The vectorised function can be written as
\begin{equation}\label{eq:alsing}
	{\bm f}({\bm x}) = \left({\bm \gamma} +\left( 1+ {\rm e}^{-{\bm \beta}\odot {\bm x}} \right)^{-1} \odot \left( 1 - {\bm \gamma}\right)\right) \odot {\bm x},
\end{equation}
such that ${\bm \beta}$ and ${\bm \gamma}$ can be trained for each node in each layer. This allows the network to find optimal configurations of the activation function tailored for the output of each individual node.

When using this activation function instead of ReLU or a sigmoid type function such as $\tanh$, it becomes abundantly clear that it leads to much better results given the same amount of training (see \cite{Nygaard:2022wri} for a more detailed discussion).

\subsection{Sampling methods for training data}\label{sec:sampling}

It is well known that the precision of emulators is seriously affected by differences in available training data density across the model parameter domain. Assuming for simplicity that the model parameter domain is given by a rectangular shape in $\mathbb{R}^n$, the simplest selection of training data points which has constant expected density is that of a latin hypercube. Indeed this choice is often made when generating the initial training data set for an emulator. However, the latin hypercube suffers greatly from the curse of dimensionality - in high dimensions the number of training data points corresponding to regions of high likelihood will be exceedingly small, and most of the emulating power of the network will be spent on emulating regions of little actual interest.

Various methods for remedying this problem have been applied in the literature and typically involve selecting training points with a density proportional to some function of the posterior. While this method does lead to a close to optimal training point selection, it suffers from the problem that the true posterior is unknown and requires running a prohibitively large number of function evaluations to compute.
However, assuming that a reasonable approximation to the true posterior can be computed using only the surrogate likelihood from a previously trained network, this can be used to sample new points in good locations in the model parameter space.
As an example, \connect{} employs such a method to iteratively train a network using new points selected from a MCMC run using the network trained on the previous iteration of points.

The method we use here is conceptually similar to the one used in \connect{}, but employs a more sophisticated scheme for deciding whether a new point should be added.

At each iteration the aim is to acquire new training data such that the density, $q(\theta)$, of training points is proportional to a tempered version of the true likelihood,
\begin{equation}
q(\theta) \propto {\cal L}^{1/T_T},
\end{equation}
with the ``training temperature'' $T_T$ a free and tunable hyper parameter.
At this stage the pre-existing training data consists of $N$ points for which the true likelihood is known
\begin{equation}
S = \{(\theta_j,{\cal L}(\theta_j)), j \in \{1,...,N\}\}.
\end{equation}
The integral of $q(\theta)$ across the input parameter domain should fulfill the condition
\begin{equation}
\int q(\theta) d\theta = N,
\end{equation}
and if we write $q(\theta) = c {\cal L}^{1/T_T}$ we find that $c = N/\int {\cal L}^{1/T_T} d\theta$. In order to relate this continuous integral to properties of the existing training point cloud a number of different approaches are possible. Here we will simply use a (very crude) approximation of estimating the local density of points at the location of point $p$ via the $k^{\rm th}$ nearest neighbour
\begin{equation}
\rho_p \sim \frac{k}{V_k} \sim r_k^{-d},
\end{equation}
where $d$ is the dimensionality of the input parameter space, $V_k$ is the $d$-dimensional hypervolume enclosing the $k$-nearest neighbours, and $r_k$ is the distance to the $k^{\rm th}$ nearest neighbour and thus the radius of a hypersphere enclosing all $k$ points. In practice, 
we find this crude estimate to yield good results and to be independent of $k$ for $5 \leq k \leq 50$.
We can now write the integral as
\begin{equation}
\int {\cal L}^{1/T_T} d\theta \sim \sum_{p \in S} \frac{{\cal L}(\theta_p)^{1/T_T}}{\rho_p},
\end{equation}
and substitute this into the equation for $c$
\begin{equation}
c \sim N\left(\sum_{p \in S} \frac{\mathcal{L}(\theta_p)^{1/T_T}}{\rho_p}\right)^{-1}.
\end{equation}
A given new point $p^*$ should only be added if the local estimated density is lower than the target density, i.e., $\rho_{p^*} < q(\theta_{p^*})=c{\cal L}(\theta_{p^*})^{1/T_T}$. Dividing both sides of this inequality with the tempered likelihood function evaluated at the point yields the acceptance criterion for the new point:
\begin{equation}
\frac{\rho_{p^*}}{\mathcal{L}(\theta_{p^*})^{1/T_T}} < c. \label{eq:acceptance}
\end{equation}
\\

\noindent We note that the evidence integral can be written as
\begin{equation}
E = \int \mathcal{L}(\theta) \pi(\theta) d\theta \sim \sum_{j=1}^{N} \mathcal{L}(\theta_j) \pi(\theta_j) V_j = \sum_{j=1}^{N} \frac{\mathcal{L}(\theta_j) \pi(\theta_j)}{\rho_j},
\end{equation}
where $\pi(\theta)$ is the prior probability density. In what follows we assume uniform priors over the sampled parameter volume, so that $\pi(\theta)$ is a constant and can be absorbed into the normalisation. 
For $T_T = 1$, $c \sim N/E$ and the selection procedure will make each point contribute roughly equally to the evidence integral.
Conversely, for $T_T \to \infty$ we have ${\cal L}^{1/T_T}  \to 1$ and hence $c \sim N/\sum_{j=1}^{N} \frac{1}{\rho_j}$ so that points are selected simply to be homogeneously spaced within the input parameter volume.

\subsection{Generating new proposals for training data}

Having established a selection criterion for whether a given new point should be added to the existing pool of training data we also need a method for generating a new set of points from which to choose.
In order to do this we employ a method essentially identical to what is used in \connect{}. Based on the existing data a network is trained and then used to run a Markov Chain Monte Carlo process. The chain then directly samples the posterior of ${\cal L}^{1/T_{\rm MCMC}}$, where $T_{\rm MCMC}$ is the temperature of the chain and is a controllable hyperparameter. 
This method for generating candidate points has the advantage of being cheap because it only requires an evaluation of the surrogate likelihood and can therefore be run to high convergence. Ensemble methods are well suited to this purpose because they can efficiently utilise the surrogate likelihood's parallelisability, which is why we employ the affine-invariant ensemble sampler \emcee{} \cite{Foreman-Mackey:2012any}. The process works as follows:
\begin{enumerate}
  \item \textbf{Prior Restriction:} Before sampling begins, the prior bounds are restricted using the $n_{\rm std}$ parameter, which defines a hypercube of size $\pm n_{\rm std} \times \sigma_i$ centered on the fiducial value for the $i^{\rm th}$ parameter, where $\sigma_i$ is a reference standard deviation. The actual bounds used for sampling are determined by taking the intersection of this hypercube with the original prior bounds---i.e., the more restrictive of the two is always chosen for each parameter. This ensures that physically motivated prior constraints are always respected while allowing tighter restrictions when desired.
  \item \textbf{Initialisation:} Starting positions for the ensemble of walkers are drawn uniformly from these effective prior bounds.
  \item \textbf{MCMC Sampling:} The sampler runs adaptively until convergence criteria are met: a minimum number of effective samples ($N_{\rm ESS} \times \max(\tau)$, where $N_{\rm ESS}$ is the effective sample size per walker and $\tau$ is the integrated autocorrelation time), stability in the autocorrelation time ($\max \left|\frac{\Delta\tau}{\tau}\right| < \delta_{\tau,\rm tol}$), or until a maximum number of steps is reached.
  \item \textbf{Resampling:} Once the MCMC run is complete, $N_{\rm cand}$ candidate points are sampled from the chain using weights $w = \mathcal{L}^{\frac{1}{T_T} - \frac{1}{T_{\rm MCMC}}}$. The target value $c$ is then computed separately for the current training set $S$ and the candidate set $S_{\rm cand}$, and we use the larger of the two values to ensure adequate sampling in the region of highest posterior density.
  \item \textbf{Candidate Evaluation:} For each candidate point $p^*$, we first evaluate the acceptance criterion in eq.~\eqref{eq:acceptance} using the surrogate likelihood $\mathcal{L}_{\rm surr}$. Only candidates that pass this initial test are then evaluated using the computationally expensive true likelihood $\mathcal{L}_{\rm true}$, and the acceptance criterion is re-evaluated. This two-stage process significantly reduces the number of expensive likelihood evaluations required.
  \item \textbf{Dynamic Update:} After each point is accepted, the target concentration $c$ is dynamically updated to reflect the addition of the new training point. This ensures that the acceptance criterion adapts as the training set grows, maintaining the desired sampling density throughout the iterative process.
\end{enumerate}
The typical configuration for this process uses $n_{\rm std} = 10$, 216 walkers (i.e.\ $8 \times N_{\rm dim}$), $N_{\rm ESS} = 50$, $\delta_{\tau,\rm tol} = 0.05$, and $N_{\rm cand} = 10000$.

\section{Metrics for testing the emulator precision}\label{sec:metrics}

Having developed the procedure for generating data and training the emulator network we now proceed to test the precision of the trained emulator when used for statistical analysis using a number of different metrics.

\subsection{KL divergence and credible intervals}

A typical usage of likelihood emulators in the context of cosmological data analysis comes in the form of Bayesian posterior distributions, typically presented as a set of 1D posteriors for model and nuisance parameters obtained by integrating over all other parameters. Therefore, an excellent benchmark for how well the emulator performs in typical tasks is to compare the set of 1D posteriors obtained using the surrogate likelihood from a similar run using the true likelihood.

The most often used benchmark for comparing two probability distributions, $P$ and $Q$,  is the Kullback-Leibler divergence \cite{Kullback:1951zyt}
\begin{equation}
D_{\rm KL}(P || Q) = \int P(\theta) \log \frac{P(\theta)}{Q(\theta)} d\theta,
\end{equation}
where integration is performed over all parameters. The numerical complexity involved in this computation is similar to the computation of the Bayesian evidence and for typical runs can involve the computation of several million individual likelihood values. While this is not overly expensive for the surrogate likelihood it is prohibitively expensive for the true likelihood. 
Furthermore, since typical parameter estimation usage involves computing 1D posterior distributions we instead use the set of 1D marginalised KL divergences, defined as
\begin{equation}
D^i_{\rm KL}(p_i || q_i) = \int p_i(\theta_i) \log \frac{p_i(\theta_i)}{q_i(\theta_i)} d\theta_i,
\end{equation}
where $p_i(\theta_i)$ and $q_i(\theta_i)$ are the 1D marginalised posterior densities estimated from MCMC samples using kernel density estimation (KDE).
This heuristic benchmark is comparatively cheap to compute and furthermore better suited to typical emulator usage.

The set of 1D marginalised KL divergences are combined into the rms KL divergence
\begin{equation}
D_{\rm KL, rms} = \left(\frac{1}{N} \sum_i D^i_{\rm KL}(p_i || q_i)^2 \right)^{1/2}, \label{eq:divKLrms}
\end{equation} 
which we will use as a simple benchmark for the precision of the emulator.

Since the computation of the marginalised KL divergence requires the computation of all 1D marginalised posteriors we automatically also recover 1D credible intervals for all parameters. These can also be used as a simple benchmark for following the precision of the emulator during the training iterations.
In order to arrive at a simple scalar expression, we define the following ``credible metric'' as a benchmark for the precision with which the emulator can recover credible intervals
\begin{equation}
\Delta^{\rm CM}_i \equiv \frac{ |\theta_i^+ - \theta_{i,0}^+ | + |\theta_i^- - \theta_{i,0}^- | }{\theta_{0,i}^+ - \theta_{0,i}^-}. \label{eq:crediblemetric}
\end{equation}
Here, $\theta_i^\pm$ and $\theta_{i,0}^\pm$ are the upper and lower credible interval values for parameter $i$ for the emulator and target likelihoods respectively.
In practice we will typically use ${\rm max}_i (\Delta^{\rm CM})$, i.e.\ the parameter with the poorest quality of emulation as the actual benchmark for credible intervals.
We will refer to $\Delta^{\rm CM}$ for the 68\% and 95\% intervals as $\Delta^{\rm CM68}$ and $\Delta^{\rm CM95}$ respectively.
Finally, note that in the case of one-sided credible intervals we use $\Delta^{\rm CM}_i \equiv \frac{ |\theta_i^\pm - \theta_{i,0}^\pm |}{  | \theta_{0,i}^\pm |}$ instead. $\pm$ then refers to the one credible interval which is defined.

\subsection{Profile likelihoods}

Another quantity often used in analysis of cosmological data is the 1D profile likelihood, which for parameter $\theta_j$ is defined as
\begin{equation}
{\cal L}_{\rm prof} (\theta_j)= \max_{\theta_i \neq \theta_j} {\cal L}(\theta_j),
\end{equation}
i.e. for a given value of parameter $\theta_j$ the profile likelihood is obtained by maximising over all other parameters. 
Unlike the marginalised posterior, the emulated profile likelihood is typically very sensitive to the emulated likelihood function at very high values close to the global best-fit (see e.g.\ \cite{Nygaard:2023cus}).
Here, we will simply use the precision of the emulator at the maximum likelihood point in parameter space as a useful and simple metric.

\section{Training convergence}\label{sec:convergence}

Once the iterative training algorithm has been set up the next question is how to stop the iterations at an appropriate time so that sufficient accuracy has been reached, but without using excessive computational resources.
The \connect{} framework uses the Gelman-Rubin statistic, $R-1$, calculated between chains from each iteration to estimate the difference between iterations and terminates the procedure once $R-1$ crosses below a certain threshold.

We employ a similar approach, taking advantage of the fact that at each iteration the \emcee{} sampler generates an ensemble of walkers that can be combined into a single flattened chain with well-defined statistical properties \cite{Foreman-Mackey:2012any}. For each iteration $i$, we discard the first 50\% of steps to avoid burn-in effects, then flatten the remaining chain and compute the mean vector $\boldsymbol{\mu}_i$ and covariance matrix $\mathbf{C}_i$ from a random subsample of $\min(10^5, N_{\rm chain})$ points, where $N_{\rm chain}$ is the size of the remaining flattened chain. 

The multivariate Gelman-Rubin statistic is then computed by comparing the statistics from consecutive iterations. Following the approach of \cite{Brooks:1998}, we compute the between-chain covariance
\begin{equation}
\mathbf{B} = \mathrm{Cov}(\boldsymbol{\mu}_{i-1}, \boldsymbol{\mu}_i),
\end{equation}
and the within-chain covariance
\begin{equation}
\mathbf{W} = \frac{\mathbf{C}_{i-1} + \mathbf{C}_i}{2}.
\end{equation}
The convergence diagnostic $R-1$ is then given by the largest eigenvalue of $\mathbf{W}^{-1/2} \mathbf{B} \mathbf{W}^{-1/2}$ (appropriately normalised). Training is terminated when $R-1$ falls below a specified threshold, typically set to $0.01$.

\section{Results}\label{sec:results}

Having finalised the set-up of the training pipeline we subsequently tested \client{} on the four separate likelihoods described in Section \ref{sec:lik-fun}. 
In each case the full training pipeline is run using 10 iterations (rather than the $R-1$ stopping criterion) in order to better visualise how the emulator performance improved with increasing iterations.
Results from \emcee{} runs for the emulator are compared to either the known analytic function for the two first likelihoods and to standard MCMC runs performed with \class{} and \montepython{}.

The reference runs for the two synthetic likelihoods were produced using the Cobaya MCMC sampler (Metropolis--Hastings), run until the Gelman--Rubin convergence criterion $R-1 < 0.01$ was satisfied, yielding effective sample sizes (ESS) of $\sim\!\!\!10{,}000$--$12{,}000$ per cosmological parameter for the Gaussian case and $\sim\!14{,}000$--$26{,}000$ for the non-Gaussian case. For the two real Planck likelihoods, the reference runs were performed using \montepython{} (Metropolis--Hastings) with 6 parallel chains and an oversampling ratio of 5 for the nuisance parameters, with \class{} as the theory code. Both runs accumulated approximately $3\times10^6$ total steps with $\sim\!6\times10^5$ accepted. The $\Lambda$CDM run achieved $R-1 < 10^{-3}$ for all sampled parameters and ESS of $\sim\!8{,}000$--$9{,}000$ per cosmological parameter. The sterile neutrino run reached a maximum $R-1 \approx 1.8\times10^{-2}$ and ESS of $\sim\!600$--$2{,}300$ per cosmological parameter.

All runs presented in this work were performed with essentially identical hyperparameter settings. The neural network architecture consisted of 5 hidden layers with 512 neurons each, using the Alsing activation function with trainable $\beta$ and $\gamma$ parameters as described in Section \ref{sec:activation}. Training was performed for up to 5000 epochs with a batch size of 128, using the MSRE loss function with $\numStdVar=3$ (see Section \ref{sec:loss}), a learning rate of $10^{-4}$, a random 10\% validation split, and early stopping with patience of 250 epochs. The initial training set consisted of 5000 points drawn from a latin hypercube restricted to $\pm 10\sigma$ around the fiducial values. For the iterative training procedure we used $T_T = 7$ for the training temperature, $k=20$ for the $k$-nearest neighbour density estimation, and 5000 candidate points per iteration. The MCMC sampling was performed using \emcee{} with $8 \times N_{\rm dim}$ walkers (i.e., 216 walkers for the Gaussian and $\Lambda$CDM cases, and 232 walkers for the non-Gaussian and $\Lambda$CDM+$(N_{\text{eff},s},m_s)$ cases), a maximum of $10^5$ steps per walker, and a burn-in of 5000 steps per walker. Sampling used $T_{\rm MCMC}=7$ and convergence criteria of $N_{\rm ESS}=50$ effective samples per walker and $\delta_{\tau,\rm tol}=0.05$ for the relative change in autocorrelation time. Additional technical parameters include a rebuild frequency of 50 candidates for the k-d tree, updating the $k$-NN radii used for density estimation, chunk size of 5000 steps for autocorrelation calculations, and autocorrelation thinning factor of 10.

\subsection{Tests on synthetic likelihoods}

\begin{figure}[tb]
    \centering
    \includegraphics[width=\textwidth]{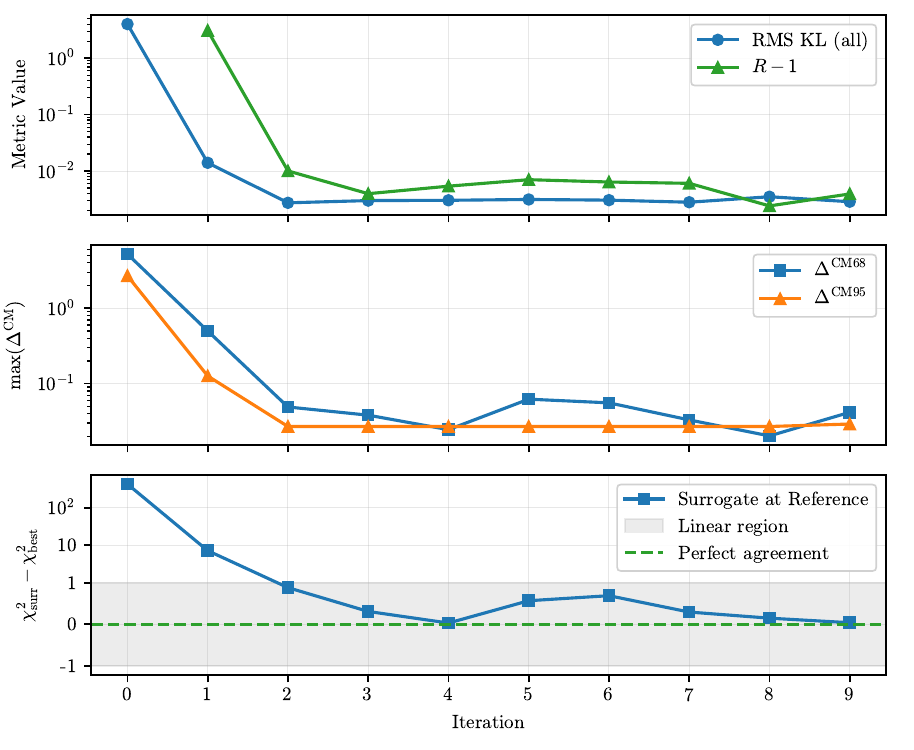}
    \caption{Convergence metrics for the synthetic Gaussian likelihood as a function of iterations. \emph{Top panel:} RMS of the KL divergence as defined in equation~\eqref{eq:divKLrms} and the Gelman–Rubin statistic of samples between iterations. \emph{Middle panel:} Maximum of the 68\% and 95\% Credible Metrics defined in equation~\eqref{eq:crediblemetric}. \emph{Bottom panel:} $\Delta \chi^2$ as evaluated by the network at the maximum likelihood point. Note that the plot is linear from $-1 < \Delta \chi^2 < 1$ and logarithmic otherwise.}
    \label{fig:rms_kl_gaussian}
\end{figure}

Figure \ref{fig:rms_kl_gaussian} shows the emulator performance as a function of iterations in the training pipeline. From the figure it is clear that already after 2 iterations the network is able to recover posterior credible intervals at a precision of around $0.05 \sigma$ and achieves a single-point precision of better than $\Delta \chi^2 \sim 1$. Running the training for 10 iterations pushes the precision somewhat further up, but it should be noted that in later iterations comparatively fewer points are added to the training data.
We also note that $R-1$ drops below 0.01 already after 2 iterations indicating that training data already samples the relevant parameter space quite well at this stage.
Finally, from the figure it can be seen that 95\% credible intervals are in general more robustly recovered than 68\% intervals because the likelihood is steeper further away from the best-fit point. This makes it easier to localise cuts in parameter space.

\begin{figure}[tb]
    \centering
    \includegraphics[width=\textwidth]{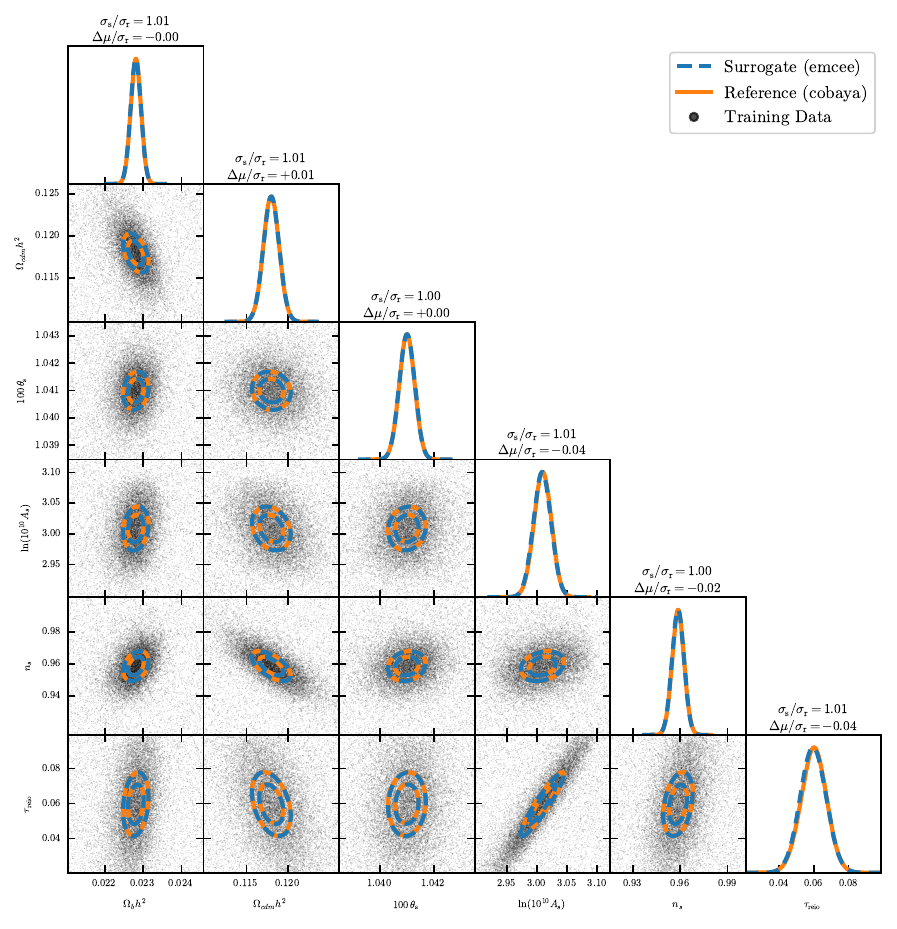}
    \caption{Triangle plot for the synthetic 27-dimensional Gaussian likelihood after the full 10 iterations. 19334 evaluations of the true likelihood with 14076 accepted points beyond the initial latin hypercube of 5000 points. Only the first 6 parameters are shown in the figure. The points show the final point cloud on which the iteration 10 network is trained. Above each marginalised posterior panel we report the error bar ratio $\sigma_s/\sigma_r$ and the normalised mean shift $\Delta\mu/\sigma_r$, where $\Delta\mu \equiv \mu_s - \mu_r$ is the difference between the surrogate and reference posterior means.}
    \label{fig:triangle_gaussian}
\end{figure}

Next, figure \ref{fig:triangle_gaussian} shows a standard triangle plot for a subset of parameters in the 27-dimensional Gaussian (named so that they match the corresponding cosmological parameters).
As can be seen from both figures, \client{} is able to emulate the target likelihood to a precision exceeding what is required for standard MCMC runs for parameter inference. Typically, a total of around $2 \times 10^4$ are used in total to train the network in iteration 10, with more than 90\% of those points having been added already by iteration 2.

\begin{figure}[tb]
    \centering
    \includegraphics[width=\textwidth]{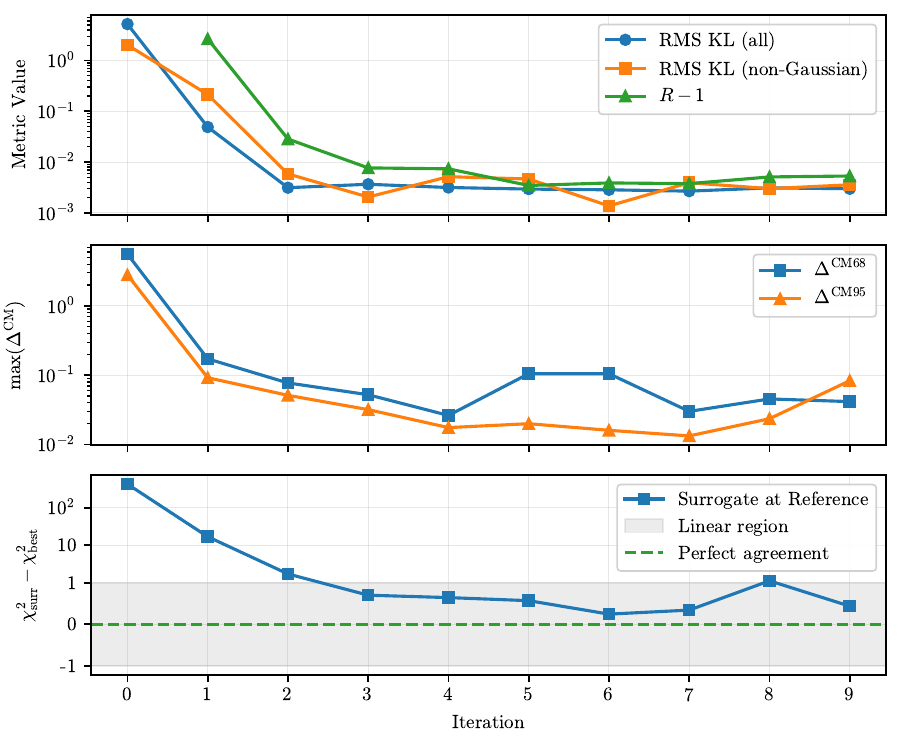}
    \caption{Convergence metrics for the synthetic likelihood that has two non-Gaussian parameters as a function of iterations. \emph{Top panel:} RMS of the KL divergence as defined in equation~\eqref{eq:divKLrms} and the Gelman–Rubin statistic of samples between iterations. \emph{Middle panel:} Maximum of the 68\% and 95\% Credible Metrics defined in equation~\eqref{eq:crediblemetric}. \emph{Bottom panel:} $\Delta \chi^2$ as evaluated by the network at the maximum likelihood point. Note that the plot is linear from $-1 < \Delta \chi^2 < 1$ and logarithmic otherwise.}
    \label{fig:rms_kl_banana}
\end{figure}

\begin{figure}[tb]
    \centering
    \includegraphics[width=\textwidth]{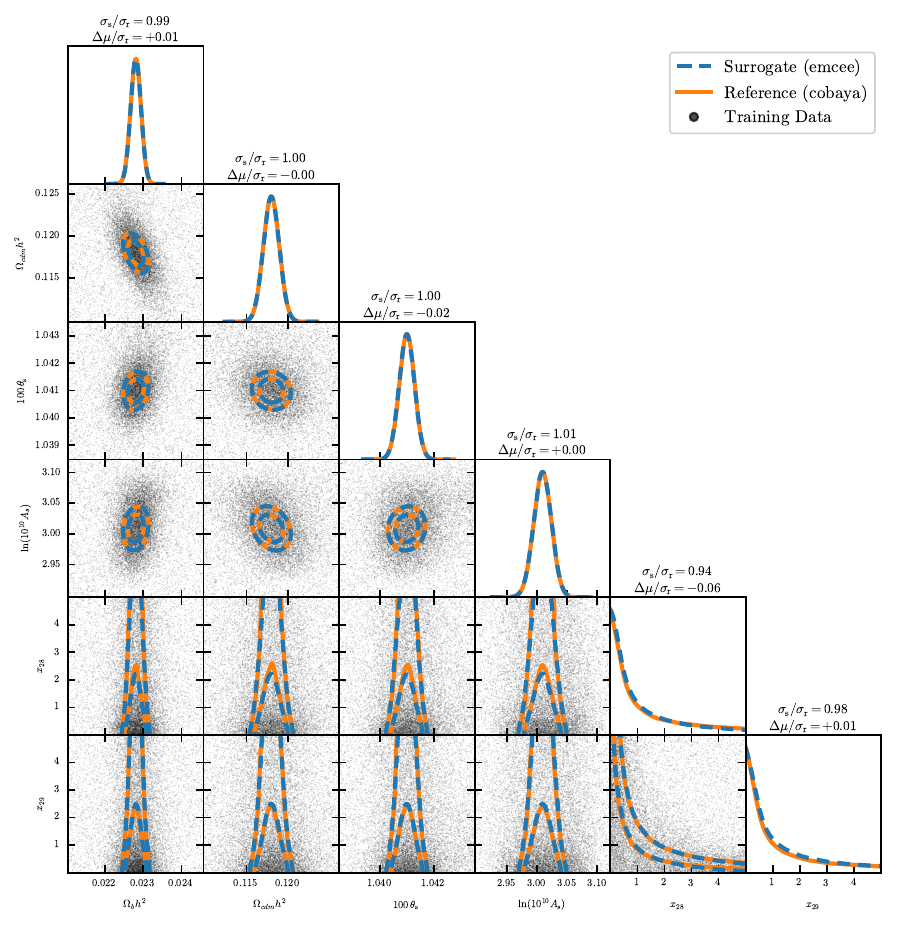}
    \caption{Triangle plot for the synthetic 29-dimensional non-Gaussian likelihood after the full 10 iterations. 17094 evaluations of the true likelihood with 11806 accepted points beyond the initial latin hypercube of 5000 points. Only parameters 1, 2, 3, 4, 28, and 29 are shown in the figure. The points show the final point cloud on which the iteration 10 network is trained. Per-parameter annotations are as described in figure~\ref{fig:triangle_gaussian}.}
    \label{fig:triangle_banana}
\end{figure}

For the 29-dimensional synthetic likelihood with two non-Gaussian directions we find essentially the same performance. However, as can be seen from figure \ref{fig:rms_kl_banana} the training takes somewhat longer to fully converge the non-Gaussian parameter directions. This happens because the required pool of training points takes longer to populate along the increasingly narrow funnel directions for the non-Gaussian directions. The $\Delta^{\rm CM}$ benchmarks are in this case also dominated by the two non-Gaussian directions.
The final result of the training is, however, comparable to the fully Gaussian likelihood.
This can also be seen from the triangle plot in figure \ref{fig:triangle_banana} in which we see excellent agreement also for the two non-Gaussian directions.

\subsection{Tests on the Planck likelihood}\label{sec:tests2}

\begin{figure}[tb]
    \centering
    \includegraphics[width=\textwidth]{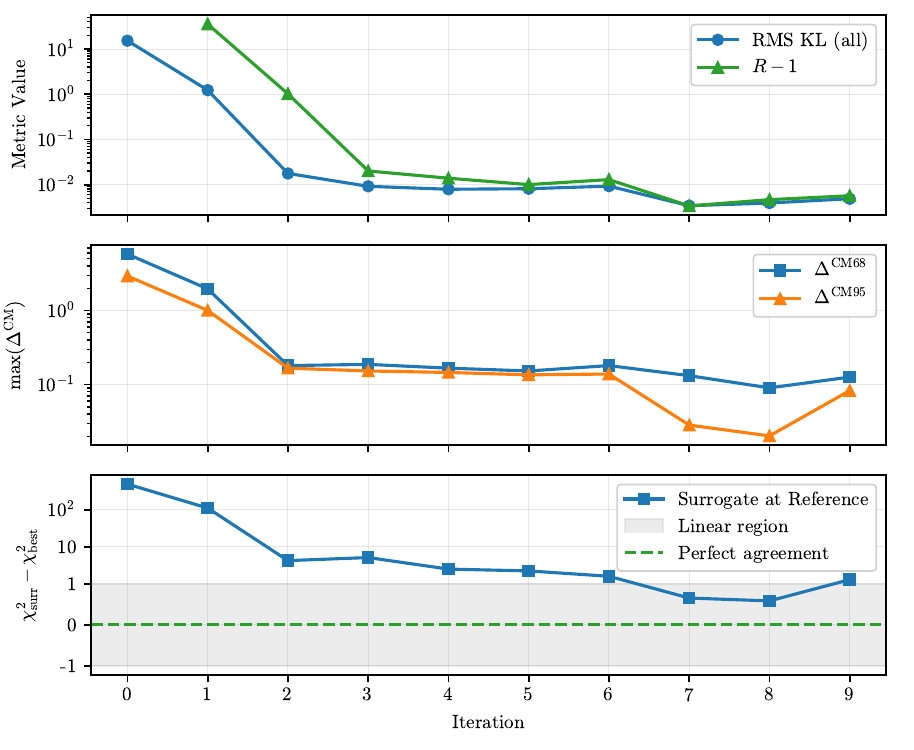}
    \caption{Convergence metrics for the Planck data with the $\Lambda$CDM model as a function of iterations. \emph{Top panel:} RMS of the KL divergence as defined in equation~\eqref{eq:divKLrms} and the Gelman–Rubin statistic of samples between iterations. \emph{Middle panel:} Maximum of the 68\% and 95\% Credible Metrics defined in equation~\eqref{eq:crediblemetric}. \emph{Bottom panel:} $\Delta \chi^2$ as evaluated by the network at the maximum likelihood point. Note that the plot is linear from $-1 < \Delta \chi^2 < 1$ and logarithmic otherwise. }
    \label{fig:rms_kl_lcdm}
\end{figure}

\begin{figure}[tb]
    \centering
    \includegraphics[width=\textwidth]{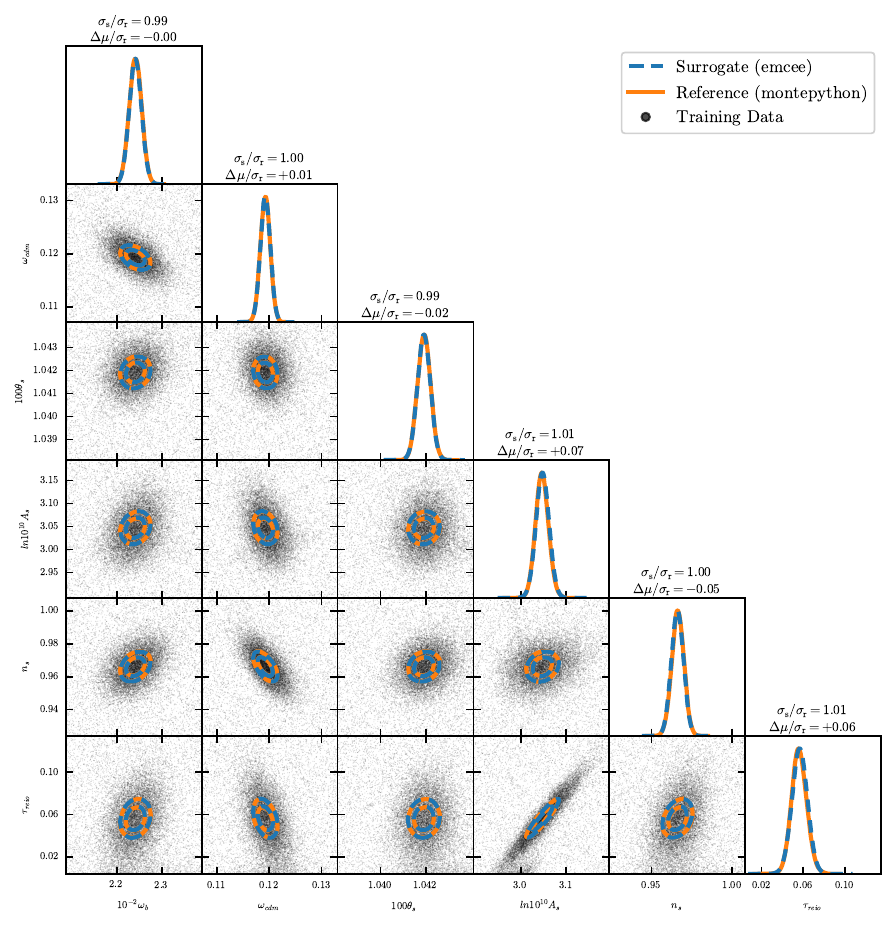}
    \caption{Triangle plot for the $\Lambda$CDM likelihood after the full 10 iterations. 22728 evaluations of the true likelihood with 12654 accepted points beyond the initial latin hypercube of 5000 points. Only the 6 cosmological parameters are shown in the figure. The points show the final point cloud on which the iteration 10 network is trained. Per-parameter annotations are as described in figure~\ref{fig:triangle_gaussian}.}
    \label{fig:triangle_lcdm}
\end{figure}

Next, we proceeded to use \client{} on actual Planck data. The hyperparameter settings for these runs were identical to those in the previous analytic likelihood tests.
From figures \ref{fig:rms_kl_lcdm} and  \ref{fig:triangle_lcdm} we see that \client{} performs almost identically well on the real Planck data with the $\Lambda$CDM model as it did with the 27-dimensional analytic likelihood. Convergence is achieved quite quickly and in terms of our metrics, the emulator again recovers posterior intervals at better than 0.1$\sigma$ after just 2 iterations. Single-point accuracy of the emulator is at around $\Delta \chi^2 \sim 2$ after two iterations, but improves to around 0.5 in later iterations - again roughly comparable to the analytic case.

\begin{figure}[tb]
    \centering
    \includegraphics[width=\textwidth]{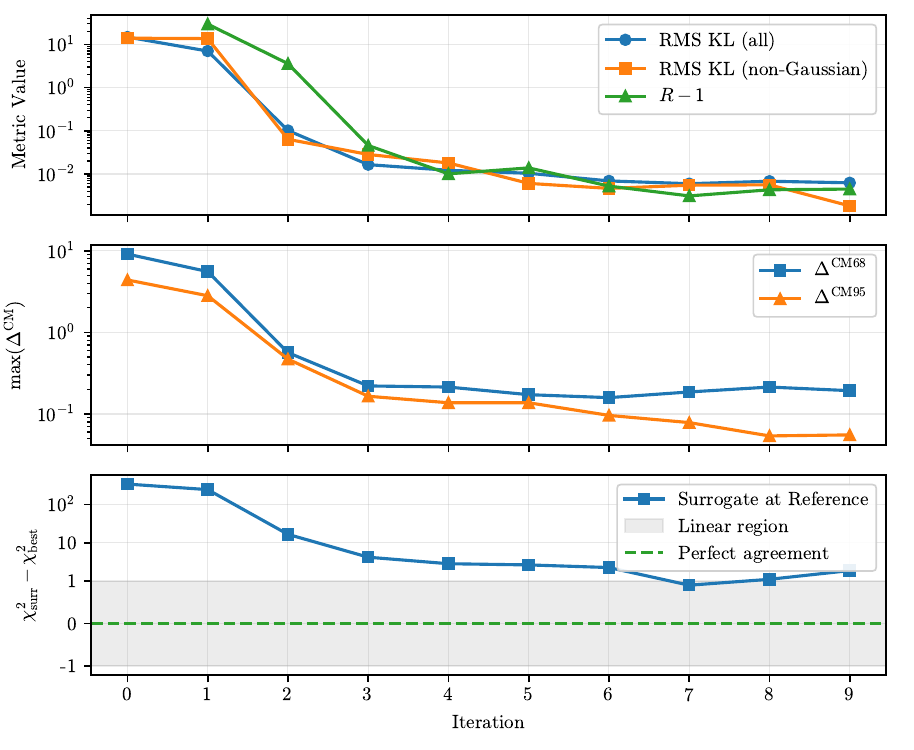}
    \caption{
    Convergence metrics for the Planck likelihood with the $\Lambda$CDM + $(N_{{\rm eff},s},m_s)$. \emph{Top panel:} RMS of the KL divergence as defined in equation~\eqref{eq:divKLrms} and the Gelman–Rubin statistic of samples between iterations. \emph{Middle panel:} Maximum of the 68\% and 95\% Credible Metrics defined in equation~\eqref{eq:crediblemetric}. \emph{Bottom panel:} $\Delta \chi^2$ as evaluated by the network at the maximum likelihood point. Note that the plot is linear from $-1 < \Delta \chi^2 < 1$ and logarithmic otherwise.}
    \label{fig:rms_kl_sterile}
\end{figure}

\begin{figure}[tb]
    \centering
    \includegraphics[width=\textwidth]{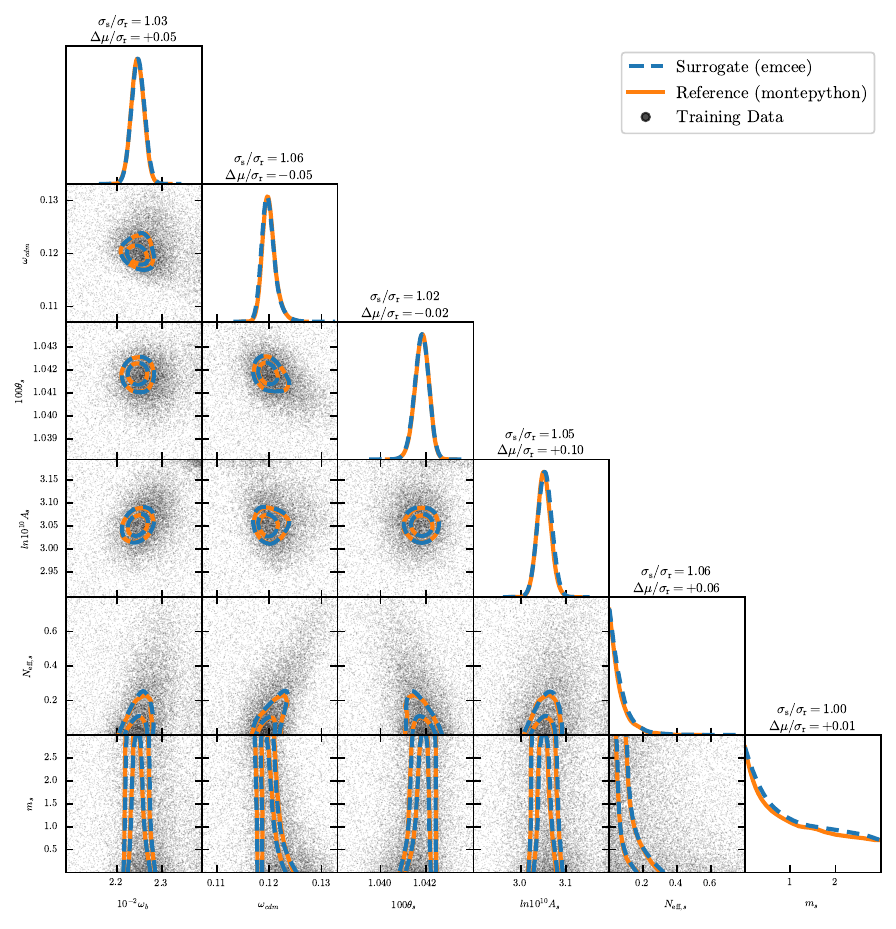}
    \caption{Triangle plot for the $\Lambda$CDM + $(N_{\text{eff},s},m_s)$ likelihood after the full 10 iterations. 22024 evaluations of the true likelihood with 11467 accepted points beyond the initial latin hypercube of 5000 points. Only 6 of the cosmological parameters are shown in the figure. The points show the final point cloud on which the iteration 10 network is trained. Per-parameter annotations are as described in figure~\ref{fig:triangle_gaussian}.}
    \label{fig:triangle_sterile}
\end{figure}

Our final case, Planck data with the $\Lambda$CDM + $(N_{\text{eff},s},m_s)$ model is shown in figures \ref{fig:rms_kl_sterile} and \ref{fig:triangle_sterile}. \client{} again achieves good emulator precision in this case, although as was also the case for the 29-dimensional analytic likelihood, convergence is slightly slower than in the more Gaussian $\Lambda$CDM case. The credible intervals do exhibit a somewhat larger deviation than in the previous cases, with the 68\% intervals only recovered at around 0.2$\sigma$ precision. However, we note here that this deviation is most likely not entirely due to lack of precision of the emulator, but rather a lack of convergence of the \class{}-based run with which we compare.
The single-point emulator precision is also somewhat worse in this case than in $\Lambda$CDM, but even here does not exceed $\Delta \chi^2 \sim 1$.
Finally, we note that we have run with the same hyperparameter settings for all four cases. In the $\Lambda$CDM + $(N_{\text{eff},s},m_s)$ this leads to slightly fewer accepted training points than for our $\Lambda$CDM case due to the increased difficulty of exploring parameter space. With more aggressive setting a larger training sample can be collected and better emulator precision achieved.

\subsection{Runtime considerations}

The iterative pipeline described in section~\ref{sec:framework} proceeds in three phases:

\begin{enumerate}
\item Evaluate likelihood on new points.
\item Train the network.
\item Sample from network and draw new points.
\end{enumerate}

We performed our runs on a node with two 32-core AMD EPYC 9334 CPU cores and two Nvidia L40S-48GB GPUs. We used 62 CPU cores and 1 of the GPUs for our runs. In phase 1 we used only CPU cores while the training in phase 2 was done on the GPU. In phase 3 we used the GPU for evaluating the network and the CPUs for the rest. With this setup, each phase of the three phases took a comparable amount of wall clock time. Specifically, for the $\Lambda$CDM-run with the full Planck likelihood, the runtime was 100 minutes. For comparison, a conventional MCMC run typically requires of order $10^6$ likelihood evaluations, which on the same hardware would correspond to approximately 27 hours --- a speed-up of roughly a factor of 16. The runtimes for the individual phases are shown in table~\ref{tab:runtimes}.

\begin{table}
\centering
\begin{tabular}{l|l}
Phase & Total runtime \\
\hline\hline
1: Evaluating likelihood on new points & 37m (22,728 points) \\
2: Training the network &  45m (1724 epochs on average) \\
3: Sampling from the network & 18m 
\end{tabular}
\caption{\label{tab:runtimes}Example of typical runtimes for a $\Lambda$CDM-run.}
\end{table}

\section{Conclusion and outlook}\label{sec:conclusion}

We have presented a novel framework for emulating cosmological likelihood functions directly rather than through the normal two-stage approach of an observable emulation followed by a likelihood evaluation.
The method was demonstrated to achieve good emulator precision on the 29D case of Planck CMB data for the sterile neutrino extension of the $\Lambda$CDM model, chosen specifically because the likelihood contains highly non-Gaussian directions in the cosmological parameter space.
For the specific setup presented here a total of less than $2 \times 10^4$ function evaluations (i.e.\ a \class{} call followed by a likelihood evaluation) call is required to achieve this precision, comparable to what is required for observable emulation frameworks such as \connect{}.
With these settings \client{} achieved a typical precision of $\Delta \chi^2 \sim 0.3$--$0.5$ for the more Gaussian cases and $\Delta \chi^2 \sim 1$ for the non-Gaussian cases in regions close to the best-fit point, and posterior credible intervals within approximately 0.1$\sigma$ of the true target credible intervals.

While \client{} was tested only on Planck data as a worked example, it works on any cosmological likelihood currently defined within the context of \montepython{} or \cobaya{} without modification. It is also worth pointing out that the \client{} framework itself is not restricted to use in cosmology, but will work on any likelihood function defined on a continuous subset of $\mathbb{R}^n$.

\section*{Reproducibility}
We have used the publicly available \client{} framework available at \url{https://github.com/AarhusCosmology/client_public}   to perform all runs presented in this work.

\section*{Acknowledgements}
We acknowledge computing resources from the Centre for Scientific Computing Aarhus (CSCAA). A.N. was supported by the Carlsberg Foundation, grant CF24-1944. We thank Marco Bonici for valuable comments on the manuscript.

\appendix

\section{Quantiles of the $\chi^2$-distribution.} \label{sec:quantile}
To relate our notion of "$\numStdVar$ sigmas" to a concrete value of $\chi^2$ we start by computing the probability integral inside $\numStdVar\sigma$ from the mean in a normal distribution. It is given by the cumulative distribution function (CDF) $\Phi(x)$ of the normal distribution as 

\begin{align}
p &= \Phi(\mu + \numStdVar\sigma) - \Phi(\mu - \numStdVar\sigma) \,,\\
& = \frac{1}{2} \left[ 1 + \text{erf} \left(\frac{\mu + \numStdVar\sigma - \mu}{\sigma \sqrt{2}} \right) \right] - \frac{1}{2} \left[ 1 + \text{erf} \left(\frac{\mu - \numStdVar\sigma - \mu}{\sigma \sqrt{2}} \right) \right] \,,\\
&= \text{erf}\left( \frac{\numStdVar}{\sqrt{2}}\right)\,.
\end{align}

We can now use the CDF of a $\chi^2$-distribution with $n$ degrees of freedom, 
\begin{align}
\text{CDF}_{\chi^2} &= \frac{1}{\Gamma\left( \frac{n}{2}\right)} \gamma \left(\frac{n}{2}, \frac{x}{2} \right) \equiv P \left(\frac{n}{2}, \frac{x}{2} \right)\,
\end{align}
where $\gamma(a,z)$ is the lower incomplete gamma-function and $P(a,z)$ is the regularised lower incomplete gamma-function. We now wish to find $x$ such that
\begin{equation}
P \left(\frac{n}{2}, \frac{x}{2} \right)= p = \text{erf}\left( \frac{\numStdVar}{\sqrt{2}} \right)\,, \label{eq:chi_squared_equation}
\end{equation}
and the solution to that equation is usually written as $P^{-1}(p)$. Specifically, we have

\begin{equation}
\chi^2(\numStdVar) = 2 P^{-1}\left(\frac{n}{2},\text{erf}\left(\frac{\numStdVar}{\sqrt{2}}\right)\right) \,. \label{eq:chi_squared_exact}
\end{equation}

Evaluating this numerically is not very stable for large values of $\numStdVar$, so we will use a simple approximation. If we define $X \equiv \chi^2$, then $(X/n)^\frac{1}{3}$ is approximately normally distributed with mean $\mu=1-\frac{2}{9n}$ and variance $\sigma^2=\frac{2}{9n}$. This transformation is known as the Wilson–Hilferty transformation~\cite{wilson1931distribution}. This means that we have the approximate value

\begin{equation}
\left(\frac{x}{n}\right)^\frac{1}{3} \simeq \mu + \numStdVar \sigma \,,
\end{equation}

if we are $\numStdVar\sigma$ away from the best-fit. Solving for $x$ gives

\begin{align}
\Delta \chi^2 = x &\simeq n (\mu + \numStdVar \sigma)^3 \,, \\
&\simeq n \left(1-\frac{2}{9n} + \numStdVar \sqrt{\frac{2}{9n}}\right)^3 \,. \label{eq:WHapprox}
\end{align}

Note, however, that because the $\chi^2$-distribution is one-sided, equation~\eqref{eq:WHapprox} is strictly speaking finding the $\numStdVar$ that solves $\Phi(\mu + \sigma) - \Phi(-\infty) = \frac{p}{2}$. Thus, we should in principle be using a modified $\numStdVar^*$ to correct for this, where

\begin{equation}
\numStdVar^* \equiv \sqrt{2} \text{erf}^{-1}\left(2 \text{erf}\left(\frac{\numStdVar}{\sqrt{2}}\right)-1\right)\,.
\end{equation}

\begin{figure}[tb]
    \centering
    \includegraphics[width=\textwidth]{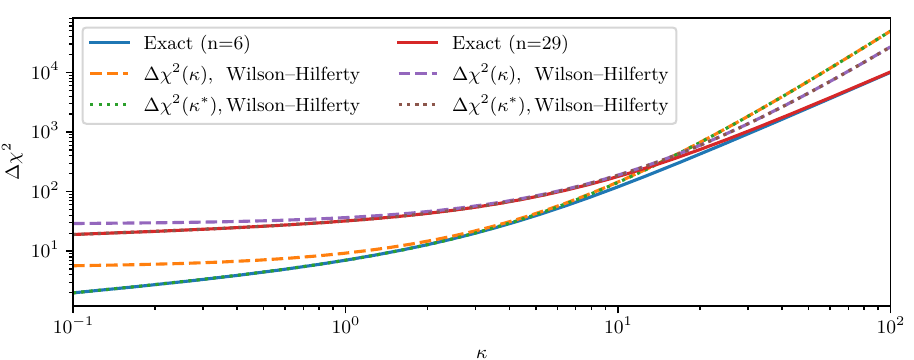}
    \caption{$\Delta \chi^2$ at $x = \numStdVar \sigma$ for $n=6$ and $n=29$. The exact calculation compared to the Wilson-Hilferty approximation in equation~\eqref{eq:WHapprox}, and the same equation with the correction $\numStdVar^*$.}
    \label{fig:delta_chi_squared_exact}
\end{figure}

In the asymptotic limit one finds $\numStdVar^* \sim \sqrt{\numStdVar^2 - 2 \log 2}$ so we have $\numStdVar^* \rightarrow \numStdVar$ for $\numStdVar\gg 1$. In figure~\ref{fig:delta_chi_squared_exact} we have shown the exact computation of $\Delta \chi^2$ compared to the Wilson–Hilferty approximation in equation~\eqref{eq:WHapprox}, with and without the correction from $\numStdVar^*$. When $\numStdVar$ is very large the approximation becomes less good since we are sampling deep in the tail of the distribution where the normality of the cubically transformed variable no longer holds. Even though the simple equation~\eqref{eq:WHapprox} breaks down in both ends, it is sufficient for our purpose because we only care about the interpretation of $\numStdVar$ in the range $1 \lesssim \numStdVar \lesssim 10$ anyway.

Lastly, evaluating equation~\eqref{eq:chi_squared_exact} numerically is non-trivial deep in the tails. Using e.g. \textsc{SciPy} we may evaluate it as

\begin{Verbatim}[commandchars=\\\{\}]
2*scipy.special.gammainccinv(n/2, scipy.special.erfc(\VarText/np.sqrt(2)))
\end{Verbatim}

but for $\numStdVar \gtrsim 38$ the complementary error function underflows in double precision. For $\numStdVar>38$ we instead asymptotically expand the defining equation~\eqref{eq:chi_squared_equation} using \url{https://dlmf.nist.gov/8.11.E2}, which leads to the equation

\begin{align}
x &= (n - 2) \log\left(\frac{x}{2}\right) + 2 \frac{n-2}{x} + A\,,\label{eq:chi_squared_fixed_point}\\
A &\equiv -2\log\Gamma\left(\frac{n}{2}\right) + \log\left( \frac{\pi \numStdVar^2}{2} \right) + \numStdVar^2 + \frac{2}{\numStdVar^2}\,,
\end{align}

for $x\equiv \Delta \chi^2$. Equation~\eqref{eq:chi_squared_fixed_point} can be solved using fixed-point iteration and converges rapidly for large $\numStdVar$.


\bibliographystyle{utcaps}
\bibliography{likelihood-emulation-2025}

\end{document}